%% file: main.tex
\documentclass{article}
\usepackage{spconf,amsmath,graphicx,bm}
\usepackage{multirow}
\usepackage{setspace}

\title{Improving Pronunciation Assessment via Ordinal Regression \\ with Anchored Reference Samples  }
\name{Bin Su$^{1,\star}$\thanks{$^\star$Work performed as intern in Microsoft}, Shaoguang Mao$^2$, Frank Soong $^{2}$, Yan Xia$^2$, Jonathan Tien$^2$, Zhiyong Wu$^1$}
\address{
$^1$Tsinghua-CUHK Joint Research Center for Media Sciences, Technologies, and Systems,\\
Shenzhen International Graduate School, Tsinghua University, Shenzhen, China\\
$^2$Microsoft Research Asia, Beijing, China\\
\small{sub18@mails.tsinghua.edu.cn, \{shamao, frankkps, yanxia, jtien\}@microsoft.com}, zywu@se.cuhk.edu.hk}
\usepackage{enumitem}
\setenumerate[1]{itemsep=0pt,partopsep=0pt,parsep=\parskip,topsep=5pt}
\begin{document}
%
\maketitle
\input{src/abstract}
\input{src/introduction}

\input{src/relatedwork}
\input{src/framework}
\input{src/experiment}
\input{src/conclusion}
\newpage
\bibliographystyle{IEEEbib}
\bibliography{strings,refs}

\end{document}

%% file: src/abstract.tex
\begin{abstract}

Sentence level pronunciation assessment is important for Computer Assisted Language Learning (CALL). Traditional speech pronunciation assessment, based on the Goodness of Pronunciation (GOP) algorithm, has some weakness in assessing a speech utterance: 1) Phoneme GOP scores cannot be easily translated into a sentence score with a simple average for effective assessment; 2) The rank ordering information has not been well exploited in GOP scoring for delivering a robust assessment and correlate well with a human rater’s evaluations. In this paper, we propose two new statistical features,  average GOP (aGOP) and confusion GOP (cGOP) and use them to train a binary classifier in Ordinal Regression with Anchored Reference Samples (ORARS). When the proposed approach is tested on Microsoft mTutor ESL Dataset, a relative improvement of Pearson correlation coefficient of 26.9\% is obtained over the conventional GOP-based one. The performance is at a human-parity level or better than human raters.
\end{abstract}

\begin{keywords}
Computer Assisted Language Learning, Ordinal Regression, Pronunciation Assessment, Goodness of Pronunciation
\end{keywords}

%% file: src/introduction.tex
\section{Introduction}
\label{sec:intro}
Sentence level pronunciation assessment is an important task in Computer Assisted Language Learning (CALL), which is commonly required by oral practice and assessment \cite{toefl,yoon2017combining_toefl}.

The Goodness of Pronunciation (GOP) \cite{wittgop} was proposed to assess pronunciation at the phoneme level. Then, the NN-based GOP \cite{hugop} replaced the GMM-based acoustic model in GOP with a deep neural network and shows a better performance. Besides, considering the change of posterior probability entropy within a phoneme, the Center GOP and transition-aware pronunciation score (TAScore) were proposed \cite{shi2020context} and achieved improved performance than GOP.

Most GOP-based algorithms e.g. \cite{wittgop,hugop,shi2020context} are designed for phoneme level pronunciation assessment. To get a pronunciation score at the sentence level, the simple average of all phoneme scores is adopted  \cite{shi2020context,li2016mispronunciationMean}. However, there are some flaws in this method: 1) simple average of all phonemes ignores the statistical difference between different phonemes.
2) The range of GOP based method is unbounded and doesn't match well with the range of human opinion scores.

Furthermore, the rank ordering information tends to be overlooked or ignored in the GOP-based algorithms, while the rank ordering information is highly related to pronunciation assessment. For example, in a five point MOS pronunciation assessment, to give a score of three, raters may focus on the intelligibility only, and to give a score of five, raters may demand more advanced pronunciation skills. When comparing two samples, how to decide which one is better is easier. By summarizing these decisions, a more reliable sentence level decision could be obtained. The pairwise sample comparison is easy to perform and a neutral way to use the rank ordering information \cite{cao2006adaptingpairwiseranking,learning2rank}.

Although the pairwise ranking algorithm is suitable for comparing two samples, the objective of pairwise ranking is to minimize the average number of inversions in ranking, which is different from the target of pronunciation assessment. To predict a sample’s rank, the pointwise ranking algorithm \cite{learning2rank} is suitable. Ordinal Regression (OR) \cite{gutierrez2015ordinal,crammer2002pranking} is a pointwise ranking algorithm and has shown its effectiveness for many applications \cite{agetask,kim2012corporatecredit,niu2016ordinalage}. 

By combining the ordinal regression with pairwise ranking algorithm, the Ordinal Regression with Anchored Reference Samples (ORARS) is proposed in speech fluency assessment \cite{mao2019nn}. The ORARS utilizes the rank ordering information to predict the rank directly. By introducing the rank ordering information, the ORARS achieves a better performance than other machine-learning or OR-based \cite{niu2016ordinalage} methods. 

In this paper, we propose a new ORARS-based framework to assess the pronunciation at the sentence level. The proposed method introduces the rank ordering information and sentence level pronunciation features to improve the assessment performance. Compared with the GOP-based method, the proposed method solves two problems: 1) The phoneme level GOP cannot be easily translated to sentence level score with simple average, which is commonly adopted by most researches, 2) The rank ordering information is ignored by the previous methods.

%% file: src/relatedwork.tex
\section{Related Work}
\subsection{Ordinal Regression}

Ordinal regression (OR) is a regression algorithm to predict a sample's rank in a set. The rank is marked with a non-negative number usually. Given a dataset $\mathbf{D}$, the $i$-th samples in $\mathbf{D}$ is $(\bm{x_i},r_i)$ where $\bm{x_i}\in\mathcal{X}$ is the input feature and $r_i\in\mathcal{R}$ is the label of the $i$-th samples in $\mathbf{D}$. The objective of OR is to find a function $r: \mathcal{X}\rightarrow \mathcal{R}$, which is similar to the regression. However, the $r_i\in \mathcal{R}$ contains the rank ordering information.

In a na\"ive approach, the ordinal regression task could be solved by regression \cite{torra2006regression4od} or classification \cite{ordinalclassification}. Besides, there are a lot of approaches to solve ordinal regression, like ordinal binary decomposition \cite{decomp1,decomp2}. However, these methods do not utilize the rank ordering information between samples.



\subsection{Ordinal Regression with Anchored Reference Samples}

The Ordinal Regression with Anchored Reference Samples (ORARS) was proposed \cite{mao2019nn} to predict the fluency score (rank). The ORARS compares each test sample with all samples in the "Anchor Set" to determine the score. It models the rank ordering information explicitly and get an improvement compared with other OR methods.

Let $\mathbf{D}$ is the training dataset, the scores of all samples in $\mathbf{D}$ are discretized to $M$ ranks firstly. Then $\mathbf{D}$ is separated into two disjoint subsets $\mathbf{D_A}$ and $\mathbf{D_T}$. $\mathbf{D_A}$ is called "Anchor Set", which contains $N$ samples in each rank. $\mathbf{D_T}$ is called training set.

A binary classifier is trained by comparing the samples $(\bm{x_i^a},y_i^a)\in\mathbf{D_A}$ and $(\bm{x_j^t},y_j^t)\in\mathbf{D_T}$ to output the probability $p_{ij}=f(\bm{x_i^a},\bm{x_j^t})$ of $y_i^a < y_j^t$. The score of a new sample $\bm{x_t}$ is predicted as Eq.\ref{eq:orars} where $p_{it}=f(\bm{x_i^a},\bm{x_t})$.

\begin{equation}
    y_t=\frac{\sum_{i=1}^{NM}p_{it}}{N}
    \label{eq:orars}
\end{equation}

%% file: src/framework.tex
\section{Proposed Framework}
\label{framework}

\begin{figure}[t]
    \centering
    \includegraphics[width=9cm]{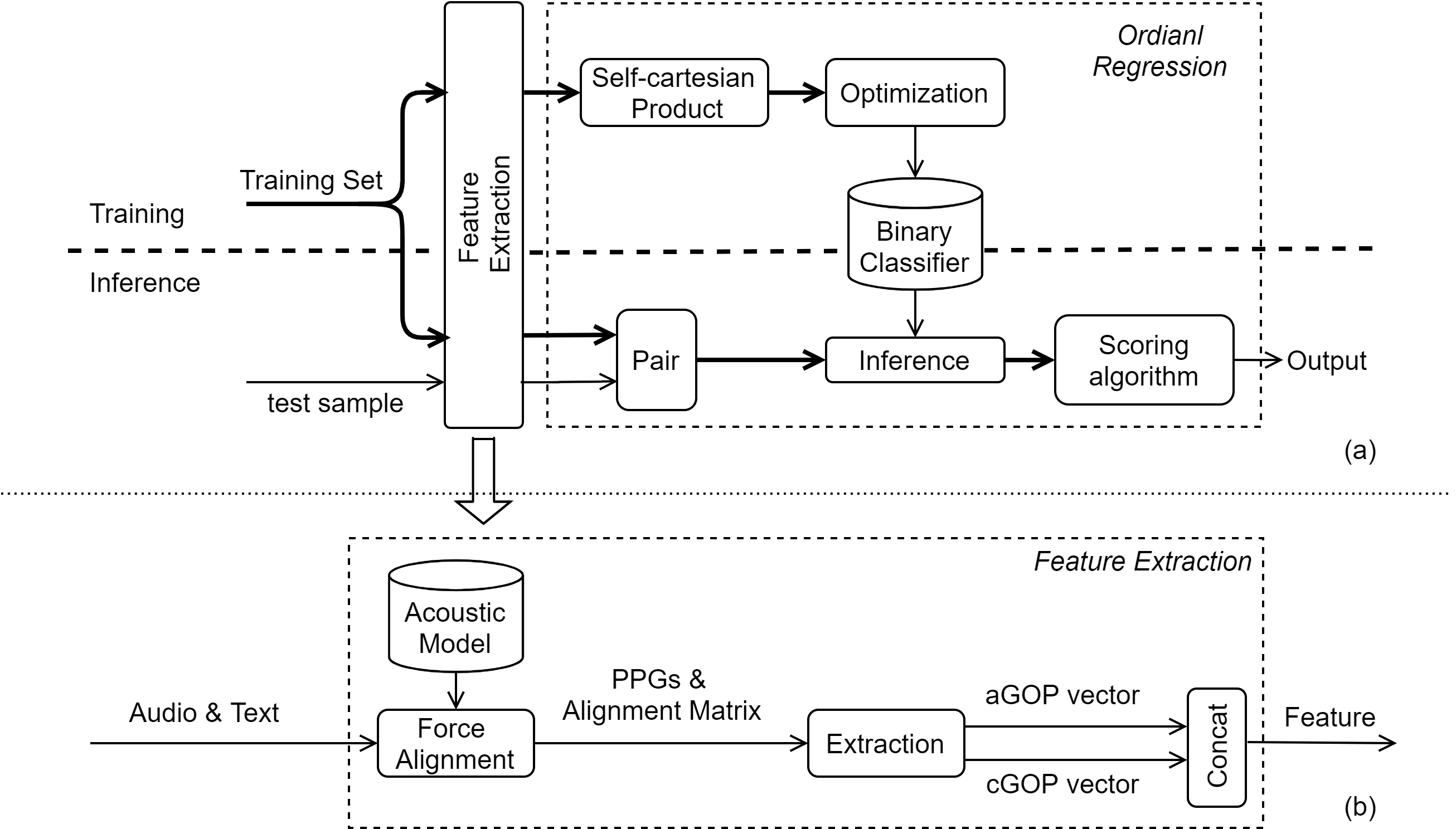}
    \caption{(a) Framework of Ordinal Regression with Anchored Reference Samples (ORARS) for Speech Pronunciation Scoring. (b) The details of Feature Extraction module.}
    \label{fig:od}
\end{figure}

The proposed framework is shown as Fig.\ref{fig:od}. The system includes two modules. The Feature Extraction (FE) module transfers raw acoustic features to fixed dimension feature vectors. The Ordinal Regression (OR) module compares two feature vectors got from FE module to predict their relative rank.

\subsection{Feature Extractor (FE) Module}
\label{sec:feature}
Average GOP (aGOP) vector and confusion GOP (cGOP) vector are two statistical features derived from the raw acoustic features. To obtain the aGOP and cGOP vectors, Phonetic Posteriorgrams (PPGs) and corresponding alignment matrix are required. The PPGs are computed from the input Mel-frequency Cepstral Coefficient (MFCC) sequences in a sentence. The alignment matrix is calculated from PPGs and text with a pre-trained acoustic model which is trained with transcribed audios.

The aGOP vector contains average GOPs on all phonemes. Let $X$ be the logarithmic PPGs with shape $(T,C)$, and $Y$ is the corresponding alignment matrix with the same shape. $T$ is the time dimension and $C$ is the phoneme dimension. Each time step in $Y$ is a one-hot vector denoted that which phoneme should be correspondingly based on the alignment. The aGOP vector $V$ could be calculated as Eq.\ref{eq:gv} which keeps the difference between different phonemes. $E_x$ is a matrix whose dimension is $(1,x)$ and all elements are $1$. The $\epsilon=10^{-6}$ is used to avoid the division by zero error. The $\frac{\ \cdot\ }{\ \cdot\ }$ indicates element-wise division. The $X^T$ means the transposition of $X$.
\begin{equation}
    V=\frac{E_CX^TY}{E_TY+\epsilon}
    \label{eq:gv}
\end{equation}

The cGOP vector indicates how pronunciation unit is confused with other phonemes. To calculate cGOP vector:
\begin{enumerate}
    \item pop the posterior probability of the target phoneme in all time steps from logarithmic PPGs $X$, the output matrix is denoted as $X_{c}$ whose shape is $(T,C-1)$;
    \item sort the elements in each time step of $X_{c}$ as descending order, the output is denoted as $X_s$;
    \item calculate the mean $M$ and standard deviation $S$ on $X_s$ along the time axis;
    \item cGOP is $[M,S]$, whose shape is $(1,2*C-2)$, where \\ $[$\ $]$ means concatenate operation. 
\end{enumerate}

\begin{figure}[t]
    \centering
    \includegraphics[width=9cm]{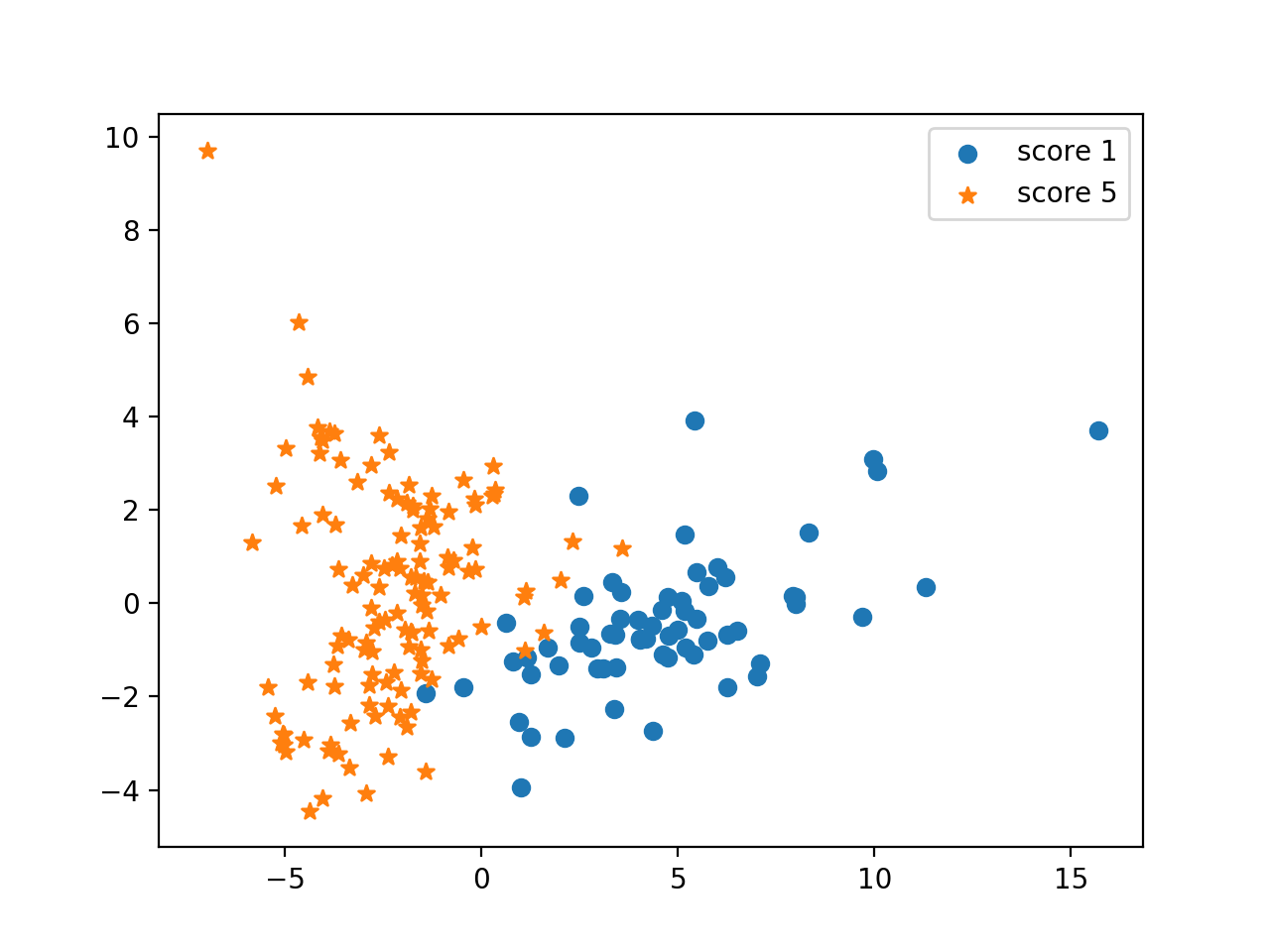}
    \caption{The scatter plot of the test utterances in their first 2 PCA axes for ``score 1" and ``score 5" classes in terms of corresponding MOS scores.}
    \label{fig:cGOP}
\end{figure}

 Fig.\ref{fig:cGOP} shows the Principal Components Analysis (PCA) \cite{pca} result of cGOP on ``score 1" and ``score 5" samples (five-mark). The scatter plot shows that the cGOP vector is quite discriminate on distinct MOS scores, and is an appropriate feature for predicting rank.

\subsection{Ordinal Regression (OR) Module}
The OR module contains a binary classifier implemented with deep neural network and  is used to predict the relative relationship in a given sample pair.

To train the binary classifier, two samples $(\bm{x_i},y_i)$ and $(\bm{x_j},y_j)$ are selected from the training dataset $\mathbf{D}$. Let $\bm{f}$ denote the FE module, the input to the binary classifier is $\bm{z_{ij}}=[\bm{f}(\bm{x_i}),\bm{f}(\bm{x_j})]$, the corresponding label $l_{ij}$ could be assigned as Eq.\ref{eq:get_label}
\begin{equation}
    l_{ij}=\left\{
    \begin{aligned}
    1\quad& y_i>y_j \\
    0\quad& y_i\le y_j 
    \end{aligned}
    \right.
    \label{eq:get_label}
\end{equation}

The output of binary classifier is $[p_{ij}^0,p_{ij}^1]$ with input $\bm{z_{ij}}$, which is activated by softmax function. The loss function could be written as Eq.\ref{eq:loss}. When the training loop is done, the $p_{ij}^1$ could be regraded as the probability of $y_i>y_j$.
\begin{equation}
    loss_{ij}=-(1-l_{ij})*logp_{ij}^0-l_{ij}*logp_{ij}^1
    \label{eq:loss}
\end{equation}

To reduce the subjective drafts from human raters, a weight function is introduced to the loss as Eq.\ref{eq:weight}. When two samples' scores are close, their loss weight will be decreased. The weight function mitigates the risk of incorrect label and makes the training process more stable. 
\begin{equation}
    w_{ij}=min(|y_i-y_j|,1)
    \label{eq:weight}
\end{equation}
\label{sec:flow}

The scoring method proposed in \cite{mao2019nn} requires a balanced "Anchor Set" which is hard to satisfy. Hence, an alternative method is proposed. The idea is that predict the rank of a sample in a given sample distribution, and then predict its score as the sample's score in corresponding rank.

Specifically, a test sample $\bm{x_t}$ is compared with all samples $(\bm{x_i},y_i)\in \mathbf{D}$ in the training set, and get a series of  posterior probabilities $\{p_{ti}^1|i\in[1,N]\}$ as mentioned above, where $N$ is the size of $\mathbf{D}$. The rank of $\bm{x_t}$ could be determined by Eq.\ref{eq:rank}
\begin{equation}
    k=\sum_{i=1}^n p_{ti}^1+1
    \label{eq:rank}
\end{equation}
Then the score in $\{y_i|i\in[1,n]\}$ with rank $\lfloor k\rfloor$ is the predicted score, where the $\lfloor$\ $\rfloor$ means round down.

%% file: src/experiment.tex
\section{Experiment}
\subsection{Datasets}
\label{sec:dataset}
The proposed model has two sub-modules needed to be trained: the acoustic model (AM) in the FE module and the binary classifier in OR module. In our experiment, two pre-trained acoustic models are used to show the robustness of proposed method, and are noted as AM1 and AM2, respectively.

The AM1 is trained with all data in WSJ0 \cite{wsj0}, LibriSpeech \cite{panayotov2015librispeech}, Common Voice \cite{commonvoice}, King ASR 249, American Children Words. The AM2 is provided by Microsoft Azure Speech Service, trained on the Microsoft EN* dataset containing more than 100k hours speech. AM2 has better performance than AM1 for it is trained with more data.

The binary classifier is trained with Microsoft mTutor ESL dataset, which contains 2445 utterances, and each utterance is rated by four raters (experienced English teachers) with $\{0,1,2,3,4,5\}$. The mean of the four scores is employed as the label. The five-fold cross-validation is used to evaluate the  model performance.


\subsection{Metrics}
The performance of different models are evaluated by Mean Absolute Error (MAE), Pearson Correlation Coefficient (PCC) and Spearman Correlation Coefficient (SCC) \cite{spearman1961general}.

Let $\bm{x}=[x_1,x_2,\cdots,x_N]$ and $\bm{y}=[y_1,y_2,\cdots,y_N]$ be two sequences. The MAE and PCC could be calculated as following
\begin{itemize}
    \item MAE
    \begin{equation}
        MAE=\frac{1}{N}\sum_{i=1}^N{|x_i-y_i|}
        \label{eq:mae}
    \end{equation}
    \item PCC
    \begin{equation}
        PCC(\bm{x},\bm{y})=\frac{cov(\bm{x},\bm{y})}{std(\bm{x})*std(\bm{y})}
    \end{equation}
    where $cov(\bm{x},\bm{y})$ donates the covariance between $\bm{x}$ and $\bm{y}$, and $std(\bm{x})$ is the standard deviation of sequence $\bm{x}$. The $*$ is the product between two scalars.
\end{itemize}

The PCC and SCC describe the correlation between the predictions and the ground truths in different perspectives. The higher the PCC or SCC, the better the assessment performance. The MAE describes the error between the predictions and ground truths. A smaller MAE denotes a better performance.

\subsection{Experiment settings}
The AM1 is a TDNN acoustic model trained with Kaldi's \cite{kaldi} default hyper-parameter settings. The AM2 is provided by Microsoft Azure Speech Service without any modifications.

The binary classifier is a fully connected neural network, which contains three ReLU activated hidden layers with hidden units are 128, 256 and 128, respectively. The output layer has two output units activated by softmax function. The loss function is introduced in section \ref{sec:flow}. The binary classifier is trained with Adam optimizer for 30 epochs with the learning rate set as $10^{-4}$, and the batch size set as 1024.

To verify the proposed feature is effective on sentence level pronunciation scoring, the GOP based methods are tested including classic GOP \cite{wittgop} and latest TAScore \cite{shi2020context}. To show the ordinal regression is effective, we trained a deep neural network-based regressor (NNR) with the proposed sentence level feature as comparative models.

To train the NNR, all samples' features are extracted by FE module denoted by $\bm{f}$. Let $(\bm{x_i},y_i)$ be the $i$-th samples in training dataset, and the NNR is trained on $\mathbf{D}_f=\{(\bm{f}(\bm{x_i}),y_i)\}$.


The NNR contains three ReLU activated hidden layers with hidden units are 128, 256 and 128, respectively. The output layer has single unit without activation function. The NNR is trained by Adam optimizer for 30 epochs with learning rate set as $10^{-4}$, the Mean Square Error (MSE) as shown as Eq.\ref{eq:mse} is selected as the loss function.
The preliminary experiments denote that training with a smaller batch size could improve training speed and performance. Hence, the batch size in the NNR model is set to 4.

\begin{equation}
    MSE(x,y)=(x-y)^2
    \label{eq:mse}
\end{equation}

All the experiments are conducted on five-fold cross-validation. In each fold, 10\% of training set was used as validation set, and the validation set for all different models are the same. When training the NNR and the proposed model, the model with the smallest loss on the validation set was used in inference stage.



\subsection{Results}
\input{src/result}

The results are shown in Table.\ref{tab:result}. Compared with the GOP-based algorithm, the FE+NNR model obtains at least 17.3\% and 16.7\% relative improvement in PCC and SCC, respectively. These results indicate that the proposed sentence level feature is much more suitable for sentence level pronunciation assessment than GOP.

Comparing with the FE+NNR model, the proposed model gets at least 2.0\%, 4.1\% and 3.5\% relative performance improvement in MAE, PCC and SCC, respectively. The proposed ORARS-based framework is effective for improving pronunciation assessment over the tested algorithms.

With the sentence level feature vector and the proposed ORARS-based framework, we have improved the performance of sentence level pronunciation assessment. The proposed model, achieves at least 26.9\% and 20.8\% relative improvement over the GOP-based methods. When a better acoustic model is available, more improvement can be achieved.

Compared with the inter-rater reliability, the proposed method shows even better performance and achieves human-parity.

%% file: src/result.tex
\begin{table}[t]
    \centering
    \caption{Performance comparison between different assessment algorithms.}
    \setlength\tabcolsep{0.8em}
    \begin{tabular}{c|c|ccc}
         \hline
         Acoustic Model& Algorithm & MAE  & PCC  & SCC \\ \hline \hline
         \multirow{4}{*}{AM1} & GOP & / &0.51 & 0.48\\ 
            & TASocre & / &  0.52 & 0.48\\
            & FE+NNR & 0.51 & 0.61 & 0.56\\
            & \textbf{Proposed} & \textbf{0.50} &\textbf{0.66} & \textbf{0.58} \\ \hline
        \multirow{4}{*}{AM2} &   GOP & /& 0.50 & 0.48  \\
            &   TAScore & / & 0.51 & 0.49  \\
            & FE+NNR & 0.45 & 0.74 & 0.67 \\
            & \textbf{Proposed} & \textbf{0.42} & \textbf{0.77} & \textbf{0.70}  \\ \hline \hline
        \multicolumn{2}{c|}{Human rating$^*$} &0.61 & 0.66 & 0.63  \\ \hline
    \end{tabular}
    \small{\leftline{$^*$Human rating is computer by inter-rater method.}}
    \label{tab:result}
\end{table}


%% file: src/conclusion.tex
\section{Conclusion}

In this study, we proposed a new set of feature vectors, aGOP and cGOP, for effective assessment of pronunciation proficiency of a given speech utterance.  Additionally, the rank ordering information in the training set, which is pre-assessed (MOS scored) by human listeners for every sentence, is also exploited to train an ordinal regression-based binary classifier via an arrangement of anchored reference samples for more performance improvement. The new set of features and the ordinal regression based binary classification, when tested on the Microsoft mTutor ESL speech database,  can improve the assessment performance measured by the Pearson correlation coefficient by 27.2\%, relatively, over the traditional GOP-based approach. In comparing with human evaluation, the new assessment achieves a performance of human-parity or better and a smaller mean absolute error.

%% file: main.bbl
\begin{thebibliography}{10}

\bibitem{toefl}
Keelan Evanini and Xinhao Wang,
\newblock ``Automated speech scoring for non-native middle school students with
  multiple task types.,''
\newblock in {\em INTERSPEECH}, 2013, pp. 2435--2439.

\bibitem{yoon2017combining_toefl}
Su-Youn Yoon and Klaus Zechner,
\newblock ``Combining human and automated scores for the improved assessment of
  non-native speech,''
\newblock {\em Speech Communication}, vol. 93, pp. 43--52, 2017.

\bibitem{wittgop}
Silke~Maren Witt et~al.,
\newblock {\em Use of speech recognition in computer-assisted language
  learning},
\newblock Ph.D. thesis, University of Cambridge Cambridge, United Kingdom,
  1999.

\bibitem{hugop}
Wenping Hu, Yao Qian, Frank~K Soong, and Yong Wang,
\newblock ``Improved mispronunciation detection with deep neural network
  trained acoustic models and transfer learning based logistic regression
  classifiers,''
\newblock {\em Speech Communication}, vol. 67, pp. 154--166, 2015.

\bibitem{shi2020context}
Jiatong Shi, Nan Huo, and Qin Jin,
\newblock ``Context-aware goodness of pronunciation for computer-assisted
  pronunciation training,''
\newblock {\em arXiv preprint arXiv:2008.08647}, 2020.

\bibitem{li2016mispronunciationMean}
Kun Li, Xiaojun Qian, and Helen Meng,
\newblock ``Mispronunciation detection and diagnosis in l2 english speech using
  multidistribution deep neural networks,''
\newblock {\em IEEE/ACM Transactions on Audio, Speech, and Language
  Processing}, vol. 25, no. 1, pp. 193--207, 2016.

\bibitem{cao2006adaptingpairwiseranking}
Yunbo Cao, Jun Xu, Tie-Yan Liu, Hang Li, Yalou Huang, and Hsiao-Wuen Hon,
\newblock ``Adapting ranking svm to document retrieval,''
\newblock in {\em SIGIR}, 2006, pp. 186--193.

\bibitem{learning2rank}
Tie-Yan Liu,
\newblock {\em Learning to rank for information retrieval},
\newblock Springer Science \& Business Media, 2011.

\bibitem{gutierrez2015ordinal}
Pedro~Antonio Guti{\'e}rrez, Maria Perez-Ortiz, Javier Sanchez-Monedero,
  Francisco Fernandez-Navarro, and Cesar Hervas-Martinez,
\newblock ``Ordinal regression methods: survey and experimental study,''
\newblock {\em IEEE Transactions on Knowledge and Data Engineering}, vol. 28,
  no. 1, pp. 127--146, 2015.

\bibitem{crammer2002pranking}
Koby Crammer and Yoram Singer,
\newblock ``Pranking with ranking,''
\newblock in {\em NIPS}, 2002, pp. 641--647.

\bibitem{agetask}
K.~{Chang}, C.~{Chen}, and Y.~{Hung},
\newblock ``Ordinal hyperplanes ranker with cost sensitivities for age
  estimation,''
\newblock in {\em CVPR}, 2011, pp. 585--592.

\bibitem{kim2012corporatecredit}
Kyoung-jae Kim and Hyunchul Ahn,
\newblock ``A corporate credit rating model using multi-class support vector
  machines with an ordinal pairwise partitioning approach,''
\newblock {\em Computers \& Operations Research}, vol. 39, no. 8, pp.
  1800--1811, 2012.

\bibitem{niu2016ordinalage}
Zhenxing Niu, Mo~Zhou, Le~Wang, Xinbo Gao, and Gang Hua,
\newblock ``Ordinal regression with multiple output cnn for age estimation,''
\newblock in {\em CVPR}, 2016, pp. 4920--4928.

\bibitem{mao2019nn}
Shaoguang Mao, Zhiyong Wu, Jingshuai Jiang, Peiyun Liu, and Frank~K Soong,
\newblock ``{NN}-based ordinal regression for assessing fluency of esl
  speech,''
\newblock in {\em ICASSP}, 2019, pp. 7420--7424.

\bibitem{torra2006regression4od}
Vicen{\c{c}} Torra, Josep Domingo-Ferrer, Josep~M Mateo-Sanz, and Michael Ng,
\newblock ``Regression for ordinal variables without underlying continuous
  variables,''
\newblock {\em Information Sciences}, vol. 176, no. 4, pp. 465--474, 2006.

\bibitem{ordinalclassification}
Alan Agresti,
\newblock {\em Analysis of ordinal categorical data}, vol. 656,
\newblock John Wiley \& Sons, 2010.

\bibitem{decomp1}
Hong Wu, Hanqing Lu, and Songde Ma,
\newblock ``A practical svm-based algorithm for ordinal regression in image
  retrieval,''
\newblock in {\em Proceedings of the eleventh ACM international conference on
  Multimedia}, 2003, pp. 612--621.

\bibitem{decomp2}
Kyoung-jae Kim and Hyunchul Ahn,
\newblock ``A corporate credit rating model using multi-class support vector
  machines with an ordinal pairwise partitioning approach,''
\newblock {\em Computers \& Operations Research}, vol. 39, no. 8, pp.
  1800--1811, 2012.

\bibitem{pca}
Karl Pearson,
\newblock ``Liii. on lines and planes of closest fit to systems of points in
  space,''
\newblock {\em The London, Edinburgh, and Dublin Philosophical Magazine and
  Journal of Science}, vol. 2, no. 11, pp. 559--572, 1901.

\bibitem{wsj0}
John~S. Garofolo, David Graff, Doug Paul, and David Pallett,
\newblock ``{CSR-I (WSJ0) Complete},'' 2016.

\bibitem{panayotov2015librispeech}
Vassil Panayotov, Guoguo Chen, Daniel Povey, and Sanjeev Khudanpur,
\newblock ``Librispeech: an asr corpus based on public domain audio books,''
\newblock in {\em ICASSP}, 2015, pp. 5206--5210.

\bibitem{commonvoice}
Rosana Ardila, Megan Branson, Kelly Davis, Michael Kohler, Josh Meyer, Michael
  Henretty, Reuben Morais, Lindsay Saunders, Francis Tyers, and Gregor Weber,
\newblock ``Common voice: A massively-multilingual speech corpus,''
\newblock in {\em LREC}, May 2020, pp. 4218--4222.

\bibitem{spearman1961general}
Charles Spearman,
\newblock ``General intelligenc objectively determined and measured.,''
\newblock 1961.

\bibitem{kaldi}
Daniel Povey, Arnab Ghoshal, Gilles Boulianne, Lukas Burget, Ondrej Glembek,
  Nagendra Goel, Mirko Hannemann, Petr Motlicek, Yanmin Qian, Petr Schwarz, Jan
  Silovsky, Georg Stemmer, and Karel Vesely,
\newblock ``The kaldi speech recognition toolkit,''
\newblock in {\em Workshop on ASRU}, Dec. 2011.

\end{thebibliography}
